\documentclass[prl,twocolumn,floatfix,
amsmath,amssymb,aps,superscriptaddress,nofootinbib]{revtex4-1}
\usepackage{dcolumn,amsmath}
\usepackage[T2A]{fontenc}
\usepackage{graphicx}
\graphicspath{ {./Images/} }
\usepackage{bm}
\usepackage{xfrac}
\usepackage{indentfirst} 
\usepackage{physics}

\usepackage{xcolor}

\usepackage{hyperref}

\usepackage{amsmath}
\usepackage{graphicx}
\usepackage{bm}
\usepackage{xcolor}
\usepackage{multirow}
\usepackage{amssymb}
\usepackage{soul}

\newcommand{\eEDM}{{\em e}EDM}
\newcommand{\ecm}{\ensuremath{e {\cdotp} {\rm cm}}}

\newcommand{\de}{d_\mathrm{e}}

\newcommand{\B}{\mathcal{B}} 
\newcommand{\E}{\mathcal{E}} 
\newcommand{\Ti}{$\mathcal{T}$}
\newcommand{\Par}{$\mathcal{P}$}

\begin{document}
\title{Revisited \Ti,\Par-odd spin-rotational Hamiltonian of HfF$^+$ for precise $e$EDM measurements}

\begin{abstract}
The current constraint on the electron electric dipole moment ($e$EDM), $|\de|<4.1\times 10^{-30}$ \ecm\ (90\% confidence), was recently established using the trapped $^{180}$Hf$^{19}$F$^+$ molecular ions in the $J=1$ rotational level of its $ ^3\Delta_1$ electronic state [T. S. Roussy, L. Caldwell, T. Wright, et al., arxiv:2212.11841]. The extensive experimental study of the HfF$^+$ cation provides detailed spectroscopy of the $\Omega-$doublet levels in the external rotating electric and magnetic fields.  We showed that previously developed theoretical approaches can fully reproduce the latest experimental data. Their justification from the first principles is very important for the examination of both modern molecular theory and possible systematic uncertainties in the interpretation of the experimental data obtained with high accuracy.
\end{abstract}

\author{Alexander N. Petrov}
\email{petrov\_an@pnpi.nrcki.ru}
\affiliation{Petersburg Nuclear Physics Institute named by B.P.\ Konstantinov of National Research Center ``Kurchatov Institute'' (NRC ``Kurchatov Institute'' - PNPI), 1 Orlova roscha mcr., Gatchina, 188300 Leningrad region, Russia}
\affiliation{Saint Petersburg State University, 7/9 Universitetskaya nab., St. Petersburg, 199034 Russia}
\homepage{http://www.qchem.pnpi.spb.ru    }

\author{Leonid V.\ Skripnikov}
\email{skripnikov\_lv@pnpi.nrcki.ru,leonidos239@gmail.com}
\affiliation{Petersburg Nuclear Physics Institute named by B.P.\ Konstantinov of National Research Center ``Kurchatov Institute'' (NRC ``Kurchatov Institute'' - PNPI), 1 Orlova roscha mcr., Gatchina, 188300 Leningrad region, Russia}
\affiliation{Saint Petersburg State University, 7/9 Universitetskaya nab., St. Petersburg, 199034 Russia}

\author{Anatoly V. Titov}
\email{titov_av@pnpi.nrcki.ru}
\affiliation{Petersburg Nuclear Physics Institute named by B.P.\ Konstantinov of National Research Center ``Kurchatov Institute'' (NRC ``Kurchatov Institute'' - PNPI), 1 Orlova roscha mcr., Gatchina, 188300 Leningrad region, Russia}
\affiliation{Saint Petersburg State University, 7/9 Universitetskaya nab., St. Petersburg, 199034 Russia}


\maketitle

\section{Introduction}
\label{Intro}
  At actual level of the experimental sensitivity the 
  measurement of a non-zero electron electric dipole moment (\eEDM, $\de$) would be a clear signature of the physics beyond the Standard model (SM) 
\cite{KL95,GFreview,Titov:06amin, Feng:2013, Safronova:18, eEDM_snowmass:2022}.
Recently the JILA group has obtained a new constraint on the electron electric dipole moment (\eEDM), $|\de|<4.1\times 10^{-30}$ \ecm\ (90\% confidence)
  \cite{newlimit1},
using the $^{180}$Hf$^{19}$F$^+$ ions
trapped by the rotating electric field. The measurements were performed on the ground rotational
level in the metastable first excited electronic $^3\Delta_1$ state. It overcame the latest ACME collaboration result obtained in 2018, $|d_e| \lesssim 1.1\cdot 10^{-29}\ e\cdot\textrm{cm}$ \cite{ACME:18}, by a factor of 2.4 and the first result $|\de|\lesssim 1.3\times 10^{-28}$ on the $^{180}$Hf$^{19}$F$^+$ ions  \cite{Cornell:2017} by a factor of about 32.

According to estimates within the Standard model, the $e$EDM value is roughly ten orders of magnitude smaller \cite{Khriplovich:97, Yamaguchi:2021}, so there is still   
  a wide
room for more sensitive experiments to search for new physics before encountering the SM background. A few experiments to search for the $e$EDM with other molecules are under preparation now, including ThF$^+$\cite{ThFp:2016}, BaF~\cite{BaF:2018}, YbF~\cite{YbF:2020} and YbOH~\cite{kozyryev2017precision, isaev2017laser}.

Considering a great potential for investigations of
various
\Ti, \Par-violating effects (\Ti\ is the time reversal, \Par\ is the space parity) on HfF$^+$ ions, it was proposed in Ref.~\cite{FDK14} to use $^{177}$Hf$^{19}$F$^+$ and $^{179}$Hf$^{19}$F$^+$ ions to measure the nuclear magnetic quadrupole moment (MQM) of $^{177}$Hf and $^{179}$Hf nuclei which have spins $I = 7/2$ and $I=9/2$, respectively. Then the \Ti,\Par-violating effects arising from the MQM and \eEDM\ 
in $^{177}$Hf$^{19}$F$^+$ and in $^{177}$Hf$^{19}$F$^+$ were studied in details in Refs.~\cite{Skripnikov:17b, Petrov:18b, Kurchavov:2020, Kurchavov2021}.
The MQM shift as a function of the external static electric field was calculated and it was shown that MQM effects can be distinguished from the \eEDM\ as MQM shift is different for different levels of hyperfine structure.
Recently, it was shown~\cite{Prosnyak:2023a} that the result~\cite{newlimit1} can be 
used to set an order of magnitude updated laboratory constraints on the axion-like-particle-mediated (ALP) \Ti, \Par-violating electron–electron and nucleus–electron interactions for a wide range of ALP masses.

 Beyond the limit on $e$EDM the experimental study of the HfF$^+$ cation provides highly accurate spectroscopy data of the $\Omega-$doublet levels in the external rotating electric and magnetic fields. Their calculation from the first principles is very important for the examination of modern molecular theory, possible systematic uncertainties and obtaining physical and chemical properties of the cation. These calculations
 for the electric and magnetic fields strengths and their rotation frequency corresponding to the latest measurement on $^{180}$Hf$^{19}$F$^+$ \cite{newlimit1}
 is the main goal of the paper.

\section{Level scheme of $^{180}$Hf$^{19}$F$^+$ for the electron EDM search}

\begin{figure}[h]
\centering
  \includegraphics[width=0.5\textwidth]{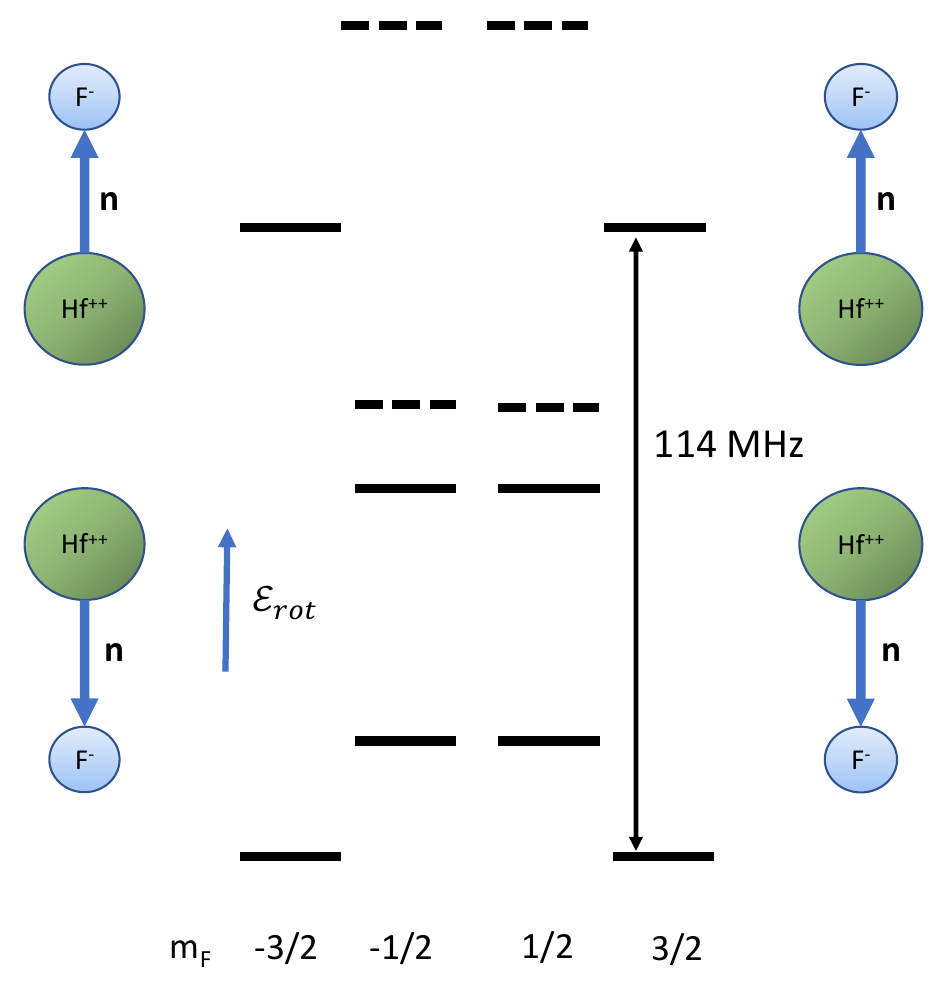}
  \caption{(Color online) Calculated hyperfine structure of ground rotational level in the metastable electronic $^3\Delta_1$ state for external rotating electric field $\E_{\rm rot}$=58 V/cm. Solid lines correspond to $F=3/2$, dashed to $F=1/2$. Energy splitting between Stark doublets for $F=3/2, m_F{=}3/2$ is about 114 MHz. Unit vector {\bf n} is directed from Hf to F. For upper (lower) Stark doublet {\bf n} is parallel (antiparallel) to the external rotating electric field.}
  \label{Hyperfine}
\end{figure}

\begin{figure}[h]
\centering
  \includegraphics[width=0.5\textwidth]{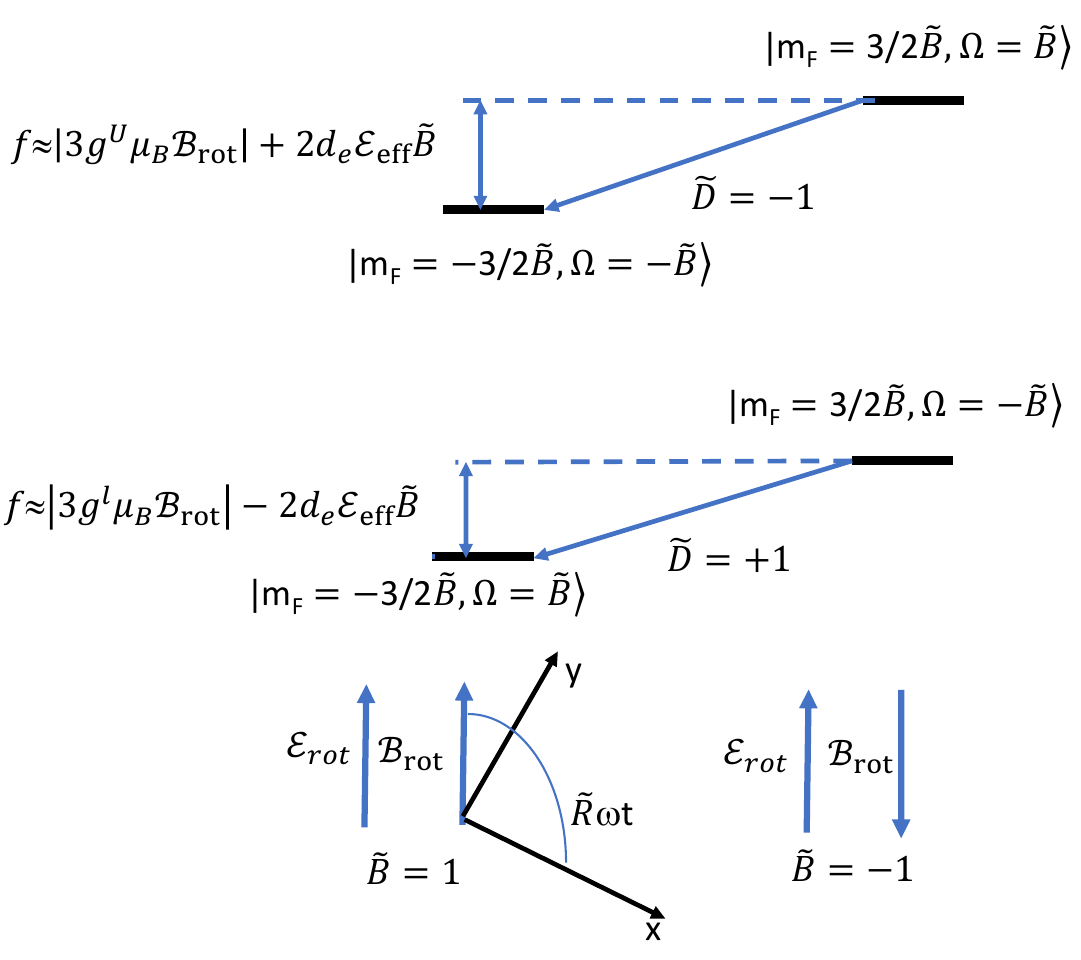}
  \caption{(Color online) The energy splitting (not to scale), $f$, between Zeeman sublevels $m_F{=}\pm3/2$. The main contribution to $f$ is from interaction (Zeeman interaction) with rotating magnetic field. Also the contribution to $f$ from \eEDM\ is indicated. External electric and magnetic field rotate in the $xy$ plane of the laboratory frame. $f$ is measured for the lower ($\tilde {\cal D}=+1$) and upper ($\tilde {\cal D}=-1$) Stark doublet,  for parallel ($\tilde {\cal B} =+1$) and antiparallel ($\tilde {\cal B} =-1$) electric and magnetic fields, for counter-clockwise ($\tilde {\cal R} = + 1$) and clockwise ($\tilde {\cal R} = -1$)  rotation of the fields around the $z$ axis.}
  \label{EDMHyperfine}
\end{figure}

 The $^{180}$Hf isotope is spinless whereas the $^{19}$F isotope has a non-zero nuclear spin $I{=}1/2$, which gives rise to the hyperfine energy splitting between the levels with total 
 (electronic-rotational-nuclear spin)
 angular momentum  $F=3/2$ and $F=1/2$, {\bf F}={\bf J}+{\bf I},
 where {\bf J} is electronic-rotational momentum.
 In the absence of external fields, each hyperfine level has two parity eigenstates known as the $\Omega$-doublet.
In the external rotating electric field 
(see Eq. (\ref{Erot}) below)
the 
$F=1/2$ states split to Stark doublets levels, whereas
$F=3/2$ state splits to two Stark doublets levels.
  One of them, with
the projection of the total momentum on the rotating field direction $m_F{=}\pm3/2$,
is of interest for the \eEDM\ search experiment. 
The corresponding energy scheme is given on Fig. \ref{Hyperfine}.
The rotating magnetic field 
(see Eq. (\ref{Brot}) below)
which is parallel or antiparallel to the rotating electric field further splits each Stark doublet to a pair of Zeeman sublevels. The energy splitting, $f$, between the sublevels $m_F{=}\pm3/2$ is measured in the experiments. 
On Fig. \Ref{EDMHyperfine} the corresponding energy scheme is given.
The measurement of $f$ is repeated under different conditions which can be characterized by binary switch parameters such as $\tilde {\cal B}$, $\tilde {\cal D}$, $\tilde {\cal R}$ being switched from $+1$ to $-1$ (see Ref. \cite{Cornell:2017, newlimit1} for details).
$\tilde {\cal B} =+1(-1)$ means that the rotating magnetic field, ${\bf B}_{\rm rot}$, is parallel (antiparallel) to the rotating electric field ${\bf E}_{\rm rot}$; $\tilde {\cal D}=+1(-1)$ means that the measurement was performed for lower (upper)
Stark level; and $\tilde {\cal R}$ defines direction for the rotation of the fields (see eqs. (\ref{Erot},\ref{Brot}) for explicit definition of $\tilde {\cal R}$). 
The measured $f$ can be represented as 
\begin{eqnarray}
\nonumber
f({\cal \tilde{D}},{\cal \tilde{B}},{\cal \tilde{R}}) = f^{0} 
+{{\cal \tilde{D}}}f^{{\cal {D}}}
+{{\cal \tilde{B}}}f^{{\cal {B}}}
+{{\cal \tilde{R}}}f^{{\cal {R}}} \\
\label{fDBR}
+{{\cal \tilde{B}}{\cal \tilde{D}}}f^{{\cal {B}}{\cal {D}}} 
+{{\cal \tilde{D}}{\cal \tilde{R}}}f^{{\cal {D}}{\cal {R}}}
+{{\cal \tilde{B}}{\cal \tilde{R}}}f^{{\cal {B}}{\cal {R}}}
+{{\cal \tilde{D}}{\cal \tilde{B}}{\cal \tilde{R}}}f^{{\cal {D}}{\cal {B}}{\cal {R}}},
\end{eqnarray}
 where notation $f^{S_1,S_2...}$ denotes a component which is odd under the switches $S_1,S_2,...$ 
 and can be calculated by formula
\begin{equation}
f^{S_1,S_2...} = \frac{1}{8}\sum_{\tilde {\cal B}, \tilde {\cal D}, \tilde {\cal R}}{S_1S_2...}f({\cal \tilde{D}},{\cal \tilde{B}},{\cal \tilde{R}}).
\label{components}
\end{equation}

The \eEDM\ signal manifests as the main contribution to $f^{ {\cal B} {\cal D}}$ channel according to 
\begin{equation}
  f^{ {\cal B} {\cal D}} = 2d_{e}E_{\rm eff},
\label{fbd}
\end{equation}
where $E_{\rm eff}$ is the effective electric field,
which can be obtained only in precise calculations of the electronic structure.
The values $E_{\rm eff}=$ 24~GV/cm \cite{Petrov:07a, Petrov:09b}, 22.5(0.9)~GV/cm \cite{Skripnikov:17c}, 22.7(1.4)~GV/cm \cite{Fleig:17} were obtained. 
According to eq. (\ref{components})
\begin{equation}
f^{{\cal {B}}{\cal {D}}} = \frac{1}{8}\sum_{\tilde {\cal B}, \tilde {\cal D}, \tilde {\cal R}}{\cal \tilde{B}}{\cal \tilde{D}}f({\cal \tilde{D}},{\cal \tilde{B}},{\cal \tilde{R}}).
\label{componentBD}
\end{equation}

Beyond the $f^{ {\cal B} {\cal D}}$ other components $f^ {0}$ (even under all switches), $f^{ {\cal D}}$, $f^ {\cal B} $ are measured with high accuracy \cite{newlimit1, newlimit2}, which, in particular, is required to control a number of  systematic effects.
As a matter of fact, all the components are measured with the same scheme but with different treatment of the raw experimental data.
In turn in Refs. \cite{Petrov:17b, Petrov:18} the precise scheme for theoretical calculation of Stark and Zeeman effects in rotating fields was developed. Recently the method was extended to the case of linear triatomic molecules \cite{Petrov:2022}.

The main goal of the paper is to calculate parameters $f^{ {\cal B} {\cal D}}$, $f^ {0}$, $f^{ {\cal D}}$ and $f^ {\cal B} $ from the \textit{first principles} and to compare them with the experimental data. 
Perfect agreement of the theoretical values with the experimental data is a very important
item for examination of both modern molecular theory and possible systematic uncertainties in interpretation of highly accurate experimental data.

\section{Theoretical methods}
Following Refs. \cite{Petrov:11,Petrov:14, Petrov:17b, Petrov:18}, the energy levels and wave functions of the  $^{180}$Hf$^{19}$F$^+$ ion are obtained by a numerical diagonalization of the molecular Hamiltonian (${\rm \bf \hat{H}}_{\rm mol}$) in the external rotating electric ${\bf E}_{\rm rot}(\rm t)$ and magnetic ${\bf B}_{\rm rot}(\rm t)$ fields 
over the basis set of the electronic-rotational wavefunctions
\begin{equation}
 \Psi_{\Omega}\theta^{J}_{M,\Omega}(\alpha,\beta)U^{\rm F}_{M_I}.
\label{basis}
\end{equation}
Here $\Psi_{\Omega}$ is the electronic wavefunction, $\theta^{J}_{M,\Omega}(\alpha,\beta)=\sqrt{(2J+1)/{4\pi}}D^{J}_{M,\Omega}(\alpha,\beta,\gamma=0)$ is the rotational wavefunction, $\alpha,\beta,\gamma$ are Euler angles, $U^{F}_{M_I}$ is the F nuclear spin wavefunctions and $M$ $(\Omega)$ is the projection of the molecule angular momentum, {\bf J}, on the lab $\hat{z}$ (internuclear $\hat{n}$) axis, $M_I=\pm1/2$ is the projection of the nuclear angular 
momentum on the same axis. Note that $M_F=M_I+M$ is not equal to $m_F$. The latter, as stated above, is the projection of the total momentum on the rotating electric field.

We write the molecular Hamiltonian for $^{180}$Hf$^{19}$F$^+$ in the form:
\begin{equation}
{\rm \bf\hat{H}}_{\rm mol} = {\rm \bf \hat{H}}_{\rm el} + {\rm \bf \hat{H}}_{\rm rot} + {\rm \bf\hat{H}}_{\rm hfs} + {\rm \bf\hat{H}}_{\rm ext}.
\end{equation} 
Here ${\rm \bf \hat{H}}_{\rm el}$ is the electronic Hamiltonian, ${\rm \bf\hat{H}}_{\rm rot}$ is the Hamiltonian of the rotation of the molecule, ${\rm \bf\hat{H}}_{\rm hfs}$ is the hyperfine interaction between electrons and fluorine nuclei as they described in Ref. \cite{Petrov:17b}  and ${\rm \bf\hat{H}}_{\rm ext}$ describes the interaction of the molecule with rotating magnetic and electric fields as it is described in Ref. \cite{Petrov:18}.

Rotating fields are expressed in terms of components that rotates in the $xy$-plane:

\begin{equation}
 {\bf E}_{\rm rot}(\rm t) =  \E_{\rm rot}(\hat{x}cos(\omega_{\rm rot}t) + \tilde {\cal R}\hat{y}sin(\omega_{\rm rot}t)),
\label{Erot}
\end{equation}

\begin{equation}
 {\bf B}_{\rm rot}(\rm t) =   \B_{\rm rot}(\hat{x}cos(\omega_{\rm rot}t) + \tilde {\cal R}\hat{y}sin(\omega_{\rm rot}t)),
\label{Brot}
\end{equation}
where $\tilde {\cal R} = \pm 1$, as described above defines direction of rotation along the $\hat{z}$ axis: ${\vec{ \omega}}_{\rm rot} = \tilde {\cal R}\omega_{\rm rot}\hat{z}$.
$\tilde {\cal R} = + 1 (-1)$ if the fields rotate counter-clockwise (clockwise) around the $\hat{z}$ axis.
Below we put $\omega_{\rm rot}/2\pi = +375$ kHz, $\E_{\rm rot}=+58$ V/cm, which are the values used in the experiment \cite{newlimit1}.
Note, that $\omega_{\rm rot}$ and $\E_{\rm rot}$ are always positive.
In this paper time-dependence of external fields
is accounted for by the transition to the rotating frame that corresponds to the first approach described in Ref. \cite{Petrov:18}.

Following Ref. \cite{Petrov:17b} we considered $^3\Delta_1$,  $^3\Delta_2$,  $^3\Pi_{0^+}$ and $^3\Pi_{0^-}$ low-lying electronic basis states.
 ${\rm \bf \hat{H}}_{\rm el}$ is diagonal on the basis set (\ref{basis}). Its eigenvalues are  transition energies of these states. They were calculated and measured in Ref.~\cite{Cossel:12}:
\begin{align}
\label{Molbasis}
\nonumber
^3\Delta_1 & : T_e=976.930~{\rm cm}^{-1}\ ,\\
\nonumber
 ^3\Delta_2 & : T_e=2149.432~{\rm cm}^{-1}\ ,\\
 \nonumber
^3\Pi_{0^-} & : T_e=10212.623~{\rm cm}^{-1}\ ,\\
 ^3\Pi_{0^+} & : T_e=10401.723~{\rm cm}^{-1}\ .
\end{align}

Electronic matrix elements required to evaluate interaction with external magnetic field (Zeeman or magnetic interaction) are~\cite{Petrov:17b}:
\begin{equation}
 \label{Gpar}
   G_\parallel = \langle ^3\Delta_1 | \hat{L}^e_{\hat{n}} -  {\rm g}_{S}\hat{S}^e_{\hat{n}}  |^3\Delta_1 \rangle, 
\end{equation}
\begin{equation}
 \label{Gperp1}
   G_{\perp}^{(1)} = \langle  ^3\Delta_1  |\hat{L}^e_- -  {\rm g}_{S}\hat{S}^e_-|^3\Delta_2  \rangle = -2.617, 
\end{equation}
\begin{equation}
 \label{Gperp2a}
   G_{\perp}^{(2a)} = \langle ^3\Delta_1  |\hat{L}^e_+ - {\rm g}_{S}\hat{S}^e_+| ^3\Pi_{0^+} \rangle  = 1.3456, 
\end{equation}
\begin{equation}
 \label{Gperp2b}
   G_{\perp}^{(2b)} = \langle ^3\Delta_1  |\hat{L}^e_+ - {\rm g}_{S}\hat{S}^e_+| ^3\Pi_{0^-} \rangle = 1.5524.  
\end{equation}
Here ${\rm g}_{S} = -2.0023$ is the free$-$electron $g$-factor, 
 ${\bf L}^e$ and ${\bf S}^e$ are the electronic orbital and electronic spin momentum operators, respectively.
 
We performed calculations for the cases when magnetic interactions with both $^3\Pi_{0^\pm}$ and $^3\Delta_2$ were taken into account and for the case when the interactions were omitted.
For the first case the body-fixed g-factor is $G_{\parallel} = 0.011768$, for the latter $G_{\parallel} = 0.012043$ and matrix elements (\ref{Gperp1}-\Ref{Gperp2b}) are set to zero. Parameters $G_{\parallel}$ were chosen in such a way that the g-factor for
$J=1$ $^3\Delta_1$ exactly corresponds to the experimental value ${\rm g} = 0.00306$ \cite{Cornell:13}.

Other electronic matrix elements for calculation of the molecular Hamiltonian were taken from Ref. \cite{Petrov:17b}, except for the hyperfine structure constant $A_{\parallel}= -62.0~{\rm MHz}$ measured in Ref. \cite{Cornell:2017} and dipole moment $D_\parallel$ for $^3\Delta_1$ which was recalculated in the present work in the accurate quantum chemical calculation (see the next section) and independently confirmed by comparison of the experimental and theoretical data.

\section{Electronic structure calculation details}

To obtain the purely \textit{ab-initio} value of the body-fixed dipole moment we used the following scheme. First we calculated the value of the dipole moment within the relativistic two-component (2c) coupled cluster method with single, double and perturbative triple cluster amplitudes, CCSD(T). The (valence) part of the generalized relativistic effective core potential (GRECP)~\cite{Mosyagin:10a,Mosyagin:16}  was employed in the electronic Hamiltonian. In the correlation calculation 52 outer-core and valence electrons were correlated, i.e. the 52e-CCSD(T) approach was employed. We used the basis set constructed in Ref.~\cite{Petrov:17b} which includes 25 $s-$, 25 $p-$, 21 $d-$, 14 $f-$, 10 $g-$, 5 $h-$ and 5 $i-$ type Gaussians for Hf and corresponds to the aug-ccpVQZ basis set \cite{Dunning:89,Kendall:92} for F which contains  6 $s-$, 5 $p-$, 4 $d-$, 3 $f-$ and 2 $g-$ contracted Gaussians and can be briefly written as (13,7,4,3,2)/[6,5,4,3,2]. The contribution of higher order correlation effects was obtained as the difference in the values of the dipole moment calculated within the coupled cluster with single, double, triple and non-iterative quadruple amplitudes, CCSDT(Q)~\cite{Kallay:6,MRCC2020}, and the CCSD(T) method. In the calculations 20 valence and outer core electrons of HfF$^+$ were correlated and the reduced basis set was used: [12,16,16,10,8]/(6,5,5,3,1)~\cite{Petrov:07a,Petrov:09b,Skripnikov:08a} basis set for Hf and [14,9,4,3]/(4,3,2,1) ANO-I basis set for F \cite{Roos:05}.
Finally, we calculated a basis set correction. For this, we turned off the spin-orbit part of the GRECP operator, i.e. switched to the scalar-relativistic approximation (for outer electrons) and calculated the correction as a difference between the values obtained within the extended basis set and the basis set used at the first step employing the coupled cluster method with single and double cluster amplitudes correlating 52 electrons.
The extended basis set for Hf contains 30 $s-$, 30 $p-$, 30 $d-$, 30 $f-$, 15 $g-$, 15 $h-$ and 15 $i-$ type functions for Hf and the uncontracted AAE4Z (19,11,6,4,2) basis set \cite{Dyall:2016} for F.
Calculations described above were performed for the equilibrium geometry of the $^3\Delta_1$ state of the HfF$^+$ cation. To obtain the value of the dipole moment for the zero vibrational level we  calculated a vibration correction as in Ref.~\cite{Petrov:17b}. 

Electronic calculations were performed within the {\sc dirac} \cite{DIRAC19,Saue:2020}, {\sc mrcc} \cite{MRCC2020,Kallay:1,Kallay:2} and {\sc cfour}~\cite{CFOUR} codes. We also employed the code developed in Refs.~\cite{Skripnikov:15b,Skripnikov:15a} to calculate  property matrix elements.

\section{Results}

The calculated value of the body-fixed dipole moment is given in Table~\ref{DzAbInitio}.
One can see that the correlation effects beyond the CCSD(T) model only modestly contribute to the value of the dipole moment. One can also see good convergence with respect to the basis set size: even a significant increase in the number of basis functions (see previous section) does not change the value of the dipole moment. The uncertainty of the final \textit{ab-initio} value of the dipole moment was calculated as a square root of squares in corrections on higher-order correlation effects, on the extended basis set and vibration correction.

\begin{table}
\caption{The calculated value of the body-fixed dipole moment $D_\parallel$ of HfF$^+$ in the $^3\Delta_1$ electronic state with the origin at the center of mass.}
\label{DzAbInitio}
\begin{tabular}{lr}
\hline
\hline
Contribution                     & $D_\parallel$, a.u. \\
\hline
52e-CCSD(T)                   & $-$1.50     \\
20e-CCSDT(Q) $-$ 20e-CCSD(T) & $-$0.02     \\
basis set correction             &  0.00     \\
vibration correction             & $-$0.01     \\
Total                            & $-$1.53(2)  \\
\hline
\hline
\end{tabular}
\end{table}

\begin{figure}[h]
\centering
  \includegraphics[width=0.5\textwidth]{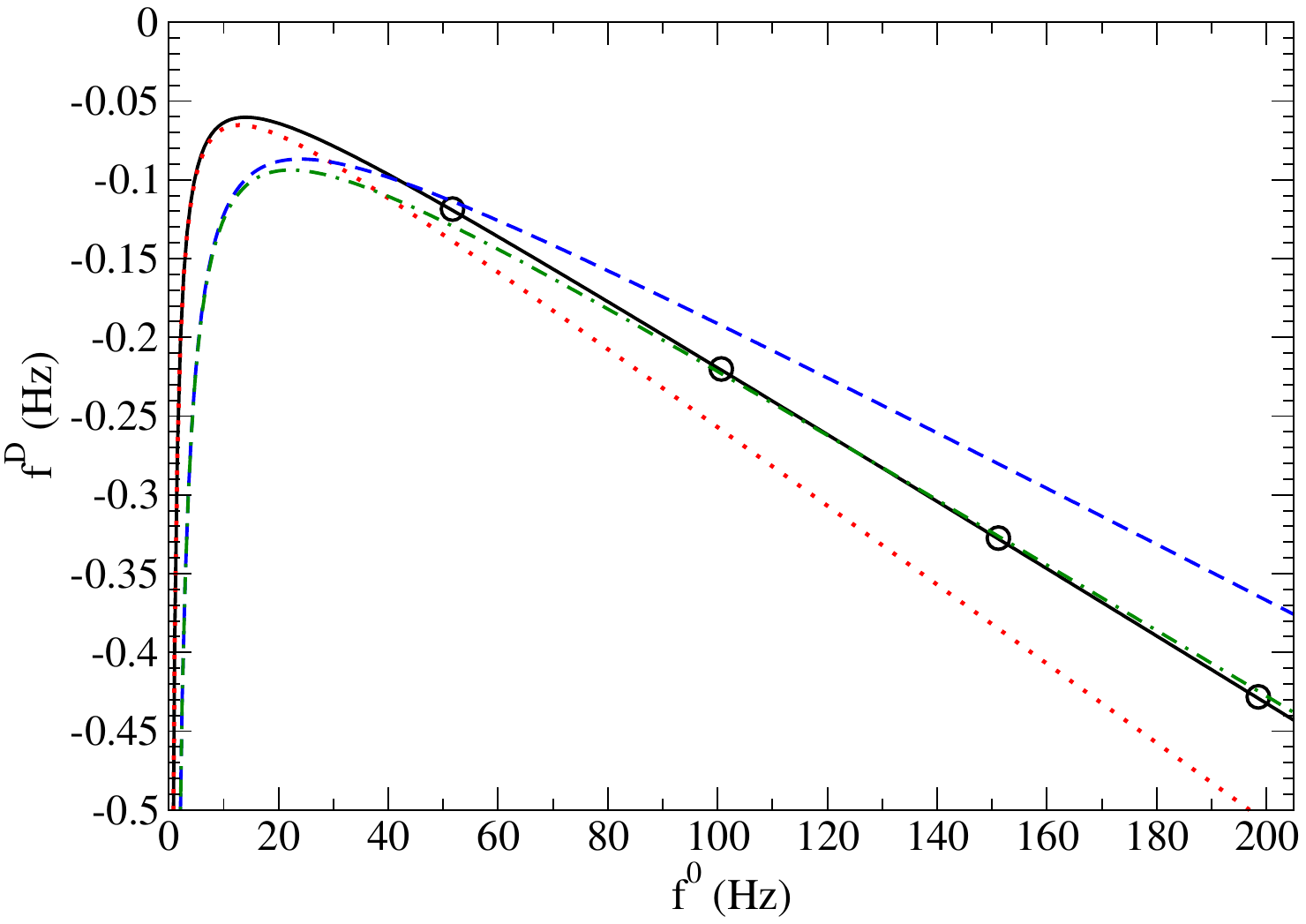}
  \caption{(Color online) $f^ {\cal D}$ as a function of $f^0$. Circles -- the experimental values \cite{newlimit2}. Solid (black) curve:  $D_\parallel=-1.53$ a.u. Dotted (red) curve:  $D_\parallel=-1.53$ a.u.,
  but interactions with both $^3\Delta_2$ and  $^3\Pi_{0^\pm}$ 
states are omitted.  Dashed (blue) curve: $D_\parallel=-1.27$ a.u. Dotted-dashed (green) curve: $D_\parallel=-1.27$ a.u., but interactions with both $^3\Delta_2$ and  $^3\Pi_{0^\pm}$  states are omitted.}
  \label{FDF0}
\end{figure}

\begin{figure}[h]
\centering
  \includegraphics[width=0.5\textwidth]{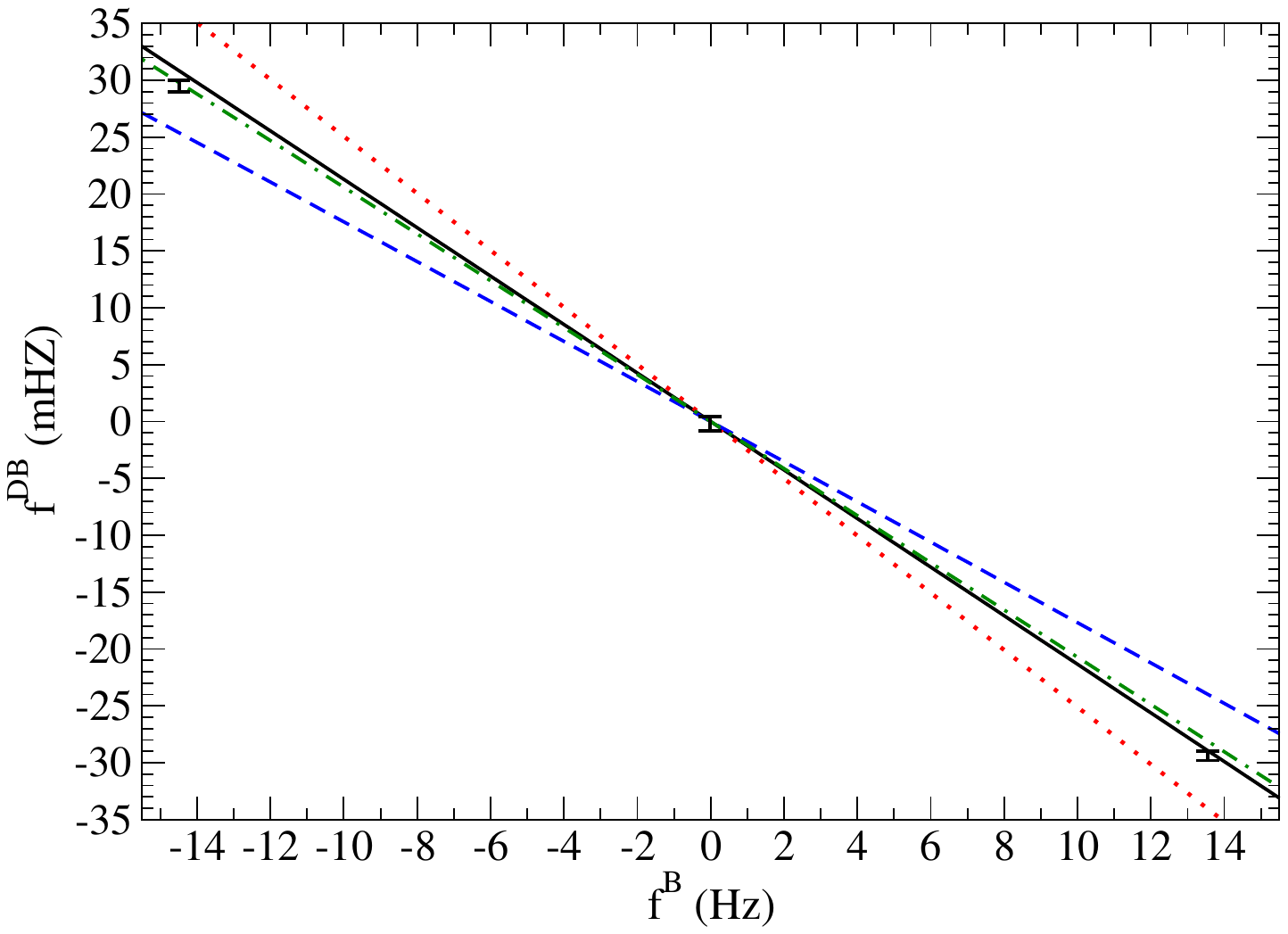}
  \caption{(Color online) $f^{ {\cal D}{\cal B}}$ as a function of $f^{ {\cal B}}$. Horizontal bands -- the experimental values, bandwidths correspond to the experimental uncertainty \cite{newlimit2}. Solid (black) curve:  $D_\parallel=-1.53$ a.u. Dotted (red) curve:  $D_\parallel=-1.53$ a.u., but interactions with both $^3\Delta_2$ and  $^3\Pi_{0^\pm}$ 
states are omitted.  Dashed (blue) curve: $D_\parallel=-1.27$ a.u. Dotted-dashed (green) curve: $D_\parallel=-1.27$ a.u., but interactions with both $^3\Delta_2$ and  $^3\Pi_{0^\pm}$  states are omitted.}
  \label{FDBFB}
\end{figure}

In Fig. \ref{FDF0} the calculated values of $f^{ {\cal D}}$ as function of $f^{ 0}$ are given. 
Fig.~\ref{FDBFB} presents the values of  $f^{ {\cal B} {\cal D}}$ as a function of $f^{ B}$.
In both Figs. the experimental values \cite{newlimit2} are given for comparison.
To plot Figs. \ref{FDF0} and \ref{FDBFB}
$f^{0}$, $f^ {\cal D}$,  $f^{ B}$, $f^{ {\cal B} {\cal D}}$ are assumed to be functions of  ${\cal B}_{\rm rot}$. For Fig.  \ref{FDBFB} the non-reversing component of the magnetic field is also added which gives the main contribution to $f^{ B}$ component. $f^0 =150.6$ Hz in Fig.  \ref{FDBFB}.

We present results for the cases when magnetic interactions with both $^3\Pi_{0^\pm}$ and $^3\Delta_2$ were taken into account and for the case when the interactions are omitted.
Calculations are also performed for different values of the 
body-fixed dipole moment, $D_\parallel$, of $^3\Delta_1$ state.
The negative value for $D_\parallel$ means that the unit vector $\hat{n}$ along the molecular axis is directed from Hf to F.
One can see that calculation that taking into account the interactions with $^3\Delta_2$,  $^3\Pi_{0^+}$ and $^3\Pi_{0^-}$ electronic states and using the dipole moment $D_\parallel = -1.53$ a.u. leads to a perfect agreement between the measured and calculated values for $f^{ {\cal D}}$ as functions of $f^{ 0}$ and a very good agreement for $f^{ {\cal B} {\cal D}}$ as a function of $f^{ B}$. As stated above $D_\parallel=-1.53$ a.u. coincides with the value calculated in Ref. \cite{Petrov:17b} (though calculated with better accuracy in this work) and 
is in good agreement with the experimental value  $D_\parallel=-1.54(1)$ a.u. \cite{Loh2013Thesis} \footnote{The same experimental group at JILA collected data sensitive to $D_\parallel$ in 2013 \cite{Loh2013Thesis}, 2014 \cite{Cossel2014Thesis, Cornell:2017} and most recently in 2022 \cite{newlimit2}, the latter shown in Figs. \ref{FDF0} and \ref{FDBFB} above. The 2013 and 2022 data are in excellent agreement with each other and with our calculations but the 2014 data is discrepant. The JILA group reports they have no explanation for the discrepant result in 2014 \cite{Cornell-Ye:2023}.}.
When the interactions with $^3\Delta_2$,  $^3\Pi_{0^+}$ and $^3\Pi_{0^-}$ states were omitted we were not able to fit all experimental data in Figs. \ref{FDF0} and \ref{FDBFB}. As an example the calculations with $D_\parallel=-1.27$ a.u. which are in good agreement with the experimental data for points $f^{ 0} = 100.67,  151.175,  198.514  $ Hz in Fig. \ref{FDF0} are given. Nevertheless, from {\it ab initio} calculations performed in this work and experiment in Ref. \cite{Loh2013Thesis} it is clear that the value $D_\parallel=-1.27$ 
is far from the real one and accounting for interactions with $^3\Pi_{0^\pm}$ and $^3\Delta_2$ is very important for an accurate calculation of $J{=}1$ levels in the electronic $^3\Delta_1$ state.

The components $f^{ {\cal D}}$ as a function of the $f^0$ and $f^{ {\cal B} {\cal D}}$ as a function of $f^{ B}$ have close relation to the g-factors of the upper, ${\rm g}^u$, and lower, ${\rm g}^l$, Stark doublets in the external {\it static} electric field. According to Refs. \cite{Cornell:2017, newlimit2}
\begin{equation}
 f^{ {\cal D}} = \frac{{\rm g}^u -{\rm g}^l} {{\rm g}^u +{\rm g}^l}  f^0 + \frac{\Delta^0\Delta^{\cal D}}{f^0},
 \label{fdf0}
\end{equation}
\begin{equation}
 f^{ {\cal B} {\cal D}} = \left( \frac{{\rm g}^u -{\rm g}^l} {{\rm g}^u +{\rm g}^l}  - \frac{\Delta^0\Delta^{\cal D}}{{f^0}^2} \right) f^{ {\cal B}},
 \label{fbdfb0}
\end{equation}
where $\Delta$ is the difference between Zeeman sublevels at zero magnetic field (which is nonzero due to the rotation of the electric field).
For the static electric field 58 V/cm and $D_\parallel=-1.53$ a.u. our calculation gives ${\rm g}^u = -3.05376\times10^{-3}$,  ${\rm g}^l = -3.06670\times10^{-3}$, $({\rm g}^u -{\rm g}^l)/ ({\rm g}^u +{\rm g}^l) = -0.002114$, $\Delta^0 = 0.7710$ Hz and $\Delta^D = -0.5501$ Hz.
If magnetic interactions with $^3\Pi_{0^\pm}$ and $^3\Delta_2$ states are not taken into account we have ${\rm g}^u = -3.05432\times10^{-3}$,  ${\rm g}^l = -3.06958\times10^{-3}$,
$({\rm g}^u -{\rm g}^l)/ ({\rm g}^u +{\rm g}^l) = -0.002491$, $\Delta^0 = 0.7709$ Hz and $\Delta^D = -0.5501$ Hz.
The calculated $f^{ {\cal D}}$ as a function of the $f^0$ and $f^{ {\cal B} {\cal D}}$ as a function of the $f^{ {\cal B}}$, given in Figs. \ref{FDF0} and \ref{FDBFB}, can be approximated with high accuracy by
\begin{equation}
 f^{ {\cal D}} = k_1  f^0 + \frac{\Delta^0\Delta^{\cal D}}{f^0}
 \label{fdf01}
\end{equation}
and
\begin{equation}
 f^{ {\cal B} {\cal D}} = k_2 f^{ {\cal B}}
 \label{fbdfb}
\end{equation}
respectively. For calculation with  $D_\parallel=-1.53$ a.u.  $k_1=-0.002151$ and $k_2 = -0.002133$ which are in good agreement with the experimental values  $k_1 = -0.002149(3)$ and $k_2 = -0.002100(20)$ obtained from $f^{ {\cal D}}$ as a function of the $f^0$ and $f^{ {\cal B} {\cal D}}$ as a function of $f^{B}$ respectively \cite{newlimit2}. 
%
The calculated values of $k_1$, $k_2$ and  $({\rm g}^u -{\rm g}^l)/ ({\rm g}^u +{\rm g}^l)$ are very close to each other but not identical as it follows from  two-level effective Hamiltonian (see Eq. (15) in Ref. \cite{newlimit2})

As it was mentioned in the Introduction the agreement between the measured and calculated values of $f^{S_1,S_2...}$ is a good test for examination of possible systematic uncertainties. For example, at the first stage of the $^{180}$Hf$^{19}$F$^+$ experiment the disagreement between calculated and measured $f^ {\cal D}$  values
as a function of $f^0$ led to a conclusion about the existence of a large (the largest, see Table II in Ref. \cite{Cornell:2017}) ``doublet population background'' systematic error. Then it was shown in Refs. \cite{Petrov:17b, Petrov:18}  that the disagreement between calculation and experiment in Ref. \cite{Cornell:2017} is on the level of interactions with $^3\Pi_{0^\pm}$ and $^3\Delta_2$ states which were not taken into account and new advanced scheme which included all the perturbations important for the $e$EDM spectroscopy was proposed. However, the previous experimental data were not accurate enough (there was only one experimental point with an error bar just on the level of the influence of the interaction with $^3\Pi_{0^\pm}$ and $^3\Delta_2$ states, see Fig. 4 in Ref. \cite{Petrov:18}) to check our method. Excellent agreement of our new calculations with new highly accurate experimental data (four points)
presented on Fig. \ref{FDF0} finally resolves the problem and declare the accurate tool for study of the systematics in the HfF$^+$ cation (and for similar systems like ThF$^+$) experiment.
Good agreement of the theory and experiment on Fig. \ref{FDBFB} points on a reliable control of systematics related with stray magnetic field.

\section{conclusion}
We calculated frequencies components $f^{0}$, $f^ {\cal D}$,  $f^{ B}$, and $f^{ {\cal B} {\cal D}}$ of the $\Omega-$double structure of $J{=}1$ rotational levels of the  $^3\Delta_1$ electronic state in the external rotating electric and magnetic fields. The high accuracy of the theoretical model introduced in Refs. \cite{Petrov:17b, Petrov:18} is demonstrated, 
 which now
 can be considered as a powerful tool helping to study systematic effects
 on HfF$^+$ ions
in experimental searches for new physics beyond the Standard model.
An accurate {\it ab~initio} value for body-frame dipole moment, $D_\parallel=-1.53(2)$ a.u., of $^3\Delta_1$ electronic state, confirmed by comparison of the calculated and experimental values for $f^{0}$, $f^ {\cal D}$,  $f^{ B}$, and $f^{ {\cal B} {\cal D}}$ is obtained.

\section{acknowledgments}
We thank Luke Caldwell, Trevor Wright, Jun Ye, and Eric Cornell for useful discussion and providing experimental data.

     Electronic structure calculations were carried out using computing resources of the federal collective usage center Complex for Simulation and Data Processing for Mega-science Facilities at National Research Centre ``Kurchatov Institute'', http://ckp.nrcki.ru/.

    $~~~$Calculations of the Stark and Zeeman effects in rotating fields were supported by the Russian Science Foundation Grant No. 18-12-00227. Calculations of property integrals were supported by the Foundation for the Advancement of Theoretical Physics and Mathematics ``BASIS'' Grant according to Project No. 21-1-2-47-1.



\end{document}